\documentclass[preprint,showpacs,preprintnumbers,tightenlines,amsmath,amssymb]{revtex4}
\usepackage{graphicx}
\usepackage{dcolumn}
\usepackage{bm}
\newcommand{\be}{\begin{equation}}
\newcommand{\ee}{\end{equation}}
\newcommand{\bee}{\begin{eqnarray}}
\newcommand{\eee}{\end{eqnarray}}
\newcommand{\eq}{\end{quote}}

\def\lsim{\displaystyle\mathop{<}_{\sim}}

\usepackage{graphicx}

\newcommand{\beq}{\begin{eqnarray}}
\newcommand{\eeq}{\end{eqnarray}}

\def\lsim{\displaystyle\mathop{<}_{\sim}}

\begin{document}      

\preprint{PNU-NTG-01/2004}
\title{Threshold production of the $\Theta^+$ in a polarized proton 
reaction}
\author{S. I. Nam}
\email{sinam@rcnp.osaka-u.ac.jp}
\affiliation{Research Center for Nuclear Physics (RCNP), Ibaraki, Osaka
567-0047, Japan}
\affiliation{Department of
Physics, Pusan National University, Pusan 609-735, Korea}

\author{A. Hosaka}
\email{hosaka@rcnp.osaka-u.ac.jp}
\affiliation{Research Center for Nuclear Physics (RCNP), Ibaraki, Osaka
567-0047, Japan}

\author{H. -Ch. Kim}
\email{hchkim@pusan.ac.kr}
\affiliation{Department of
Physics, Pusan National University, Pusan 609-735, Korea}
\date{Januray, 2004}
\begin{abstract}
We compute cross sections of $\Theta^+$ production near threshold for a
polarized proton reaction, 
$\vec p \vec p \to \Sigma^+ \Theta^+$ which was recently proposed to 
determine unambiguously the parity of $\Theta^+$.  
Within theoretical uncertainties cross sections for the allowed 
spin configuration are estimated; it is 
of order of one microbarn for the positive parity $\Theta^+$ and 
about 1/10 microbarn for the negative  parity $\Theta^+$
in the threshold energy region 
where the s-wave component dominates.  
\end{abstract}



\maketitle

The discovery of the 
$\Theta^+$~\cite{Nakano:2003qx,Barmin:2003vv,Stepanyan:2003qr,Kubarovsky:2003fi}
has triggered a tremendous amount of research  
activities both from theories and experiments~\cite{jlab}.
Hadron physics has now experienced a new stage of development
with unexpected richness.  
The surprise came not only with its relatively low mass but also 
with a very narrow width, though for the latter only the upper 
limit is known so far.  
This feature is also shared by the recently observed  
$\Xi$ states~\cite{NA49}.   
Quantum numbers such as spin and parity are not yet known neither.  
Since the parity reflects the internal dynamics of hadrons, 
it is crucially important to determine it by experiment and 
to understand by theory.  
The present theoretical situation, however, is not settled; 
chiral theories including the pioneering chiral 
soliton models~\cite{Diakonov:1997,Stancu:2003if,Hosaka:2003jv}, 
and the diquark model~\cite{Jaffe:2003sg}
predict positive parity, while recent 
lattice~\cite{Csikor,sasaki} and sum rule 
calculations~\cite{Zhu:2003ba,sugiyama} are on the other side.  

Very recently,
an unambiguous method was proposed in order to determine the parity
of the $\Theta^+$ using the reaction~\cite{Thomas:2003ak}
\beq
\vec p + \vec p \to \Theta^+ + \Sigma^+ \; \; \; 
{\rm near}\; {\rm threshold}.  
\label{reaction}
\eeq
This reaction has been previously considered for
the production of $\Theta^+$~\cite{Polyakov:1999da}, but
it has turned out that it does more for the determination of the parity,
in contrast with
a number of recent attempts using other reactions which
needed particular production mechanism~\cite{theory}.
In order to extract information of parity from (\ref{reaction}),
the only requirement is that the final state is dominated by
the s-wave component.
The s-wave dominance in the final state is then combined with the
Fermi statistics of the initial two protons and conservations of
the strong interaction, establishing the selection rule:
{\em If the parity of $\Theta^+$ is positive, the reaction
(\ref{reaction}) is allowed at the threshold region
only when the spin of the two protons is
$S = 0$, while, if it is negative the reaction is allowed only when
$S=1$.}
This situation is similar to what was used in determining the parity
of the pion.
Experimentally, the pure $S = 0$ state may not be easy to set up. 
However,
an
appropriate combination of spin polarized quantities allows to extract
information of $S = 0$ state.

\begin{figure}
\begin{center}
\includegraphics[width=7cm]{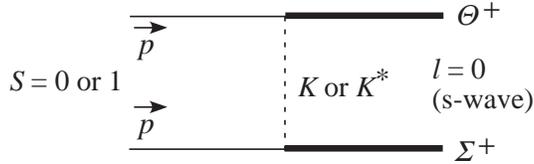}
\begin{minipage}{12cm}
\caption{\small 
Born diagrams for $\vec{p}\vec{p} \to \Theta^+ \Sigma+$.
\label{ppTS}}0
\end{minipage}
\end{center}
\end{figure}

In this letter, we perform calculations for production 
cross sections of (\ref{reaction}).  
Our purposes are: 
\begin{enumerate}

\item To check that the production reaction is indeed dominated by 
the s-wave (in other word, there is no accidental vanishing 
of s-wave contributions to invalidate the above selection rule).  

\item To estimate production cross sections within the present 
knowledge of theoretical models.
\end{enumerate}

In order to estimate the production rate, we calculate the 
Born diagrams 
of pseudoscalar kaon ($K(498)$) and vector $K^*$ ($K^*(892)$) 
exchanges, 
which are minimally needed for the present reaction (Fig.~\ref{ppTS}). 
Assuming that the parity of $\Theta^+$ is positive, 
we can take effective interaction lagrangians as follows:
\beq
\mathcal{L}_{KN\Theta}
&=&
ig_{KN\Theta}\bar{\Theta}\gamma_{5}KN 
+ {\rm (h.c.)}\, , 
\label{knt}
\\
\mathcal{L}_{KN\Sigma}
&=&
ig_{KN\Sigma}\bar{\Sigma}\gamma_{5}KN 
+ {\rm (h.c.)}\, , 
\label{kns}
\\
\mathcal{L}_{K^*N\Theta}
&=&
- g_{K^*N\Theta} \bar{\Theta}\gamma^{\mu}K^*_{\mu}N
+\frac{g^{T}_{K^*N\Theta}}{M_{\Theta}+M_{N}}
\bar{\Theta}\sigma^{\mu\nu}\partial_{\mu}K^*_{\nu}N 
+ {\rm (h.c.)}\, ,
\label{vnt}\\
\mathcal{L}_{K^*N\Sigma}
&=&
-g_{K^*N\Sigma}\bar{\Sigma}\gamma^{\mu}K^*_{\mu}N
+\frac{g^{T}_{K^*N\Sigma}}{M_{\Sigma}+M_{N}}
\bar{\Sigma}\sigma^{\mu\nu}\partial_{\mu}K^*_{\nu}N 
+ {\rm (h.c.)}\, .
\label{vnS}
\eeq
with standard notations.  
If the parity of $\Theta^+$ is negative, $\gamma_5$ matrix in 
(\ref{knt}) should be removed and in (\ref{vnt}) $\gamma_5$ should be
inserted.  
For the coupling terms of $\Sigma^+$, we employ 
the values estimated from the previous analysis; 
$g_{KN\Sigma} = 3.54$, 
$g_{K^*N\Sigma} = -2.46$ and $g^{T}_{K^*N\Sigma}$ = 1.15~\cite{stokes}.  
Since the couplings to the $\Theta^+$ is not known, 
we investigate several cases with different parameter
values.  
For $g_{KN\Theta}$ we mostly employ $g_{KN\Theta} = 3.78$, which 
is fixed by 
$\Gamma_{\Theta^+ \to KN} = 15$ MeV.  
For each case, we employ for the unknown vector $K^*$ couplings, 
$|g_{K^*N\Theta}| = |g_{KN\Theta}|/2$, as suggested by Ref.~\cite{Liu:2003hu}.
The tensor couplings are then varied within 
$|g_{K^*N\Theta}^T| \leq 2|g_{K^*N\Theta}| = |g_{KN\Theta}|$
in order to see model dependence of this process.  
As for the form factor, we employ the following form of the monopole type:
\beq
F(q^2) = \frac{\Lambda^2 - m^2}{\Lambda^2 - q^2} \, ,
\label{ff}
\eeq
where $q^2$ is the four momentum square 
and $m$ the mass of the exchanged particle (either $K$ or $K^*$).  
The cut off parameter $\Lambda$ is chosen to be 
$\Lambda = 1$ GeV. In Ref.~\cite{Oh:2003gj} the authors employed
a different type of form factor. However, the monopole type is more
often used for meson-baryon vertices. In any events, the main points
in the following discussions will not be changed by the use of
different form factors.    

\begin{figure}[tbh]
\begin{center}
\includegraphics[width=6cm]{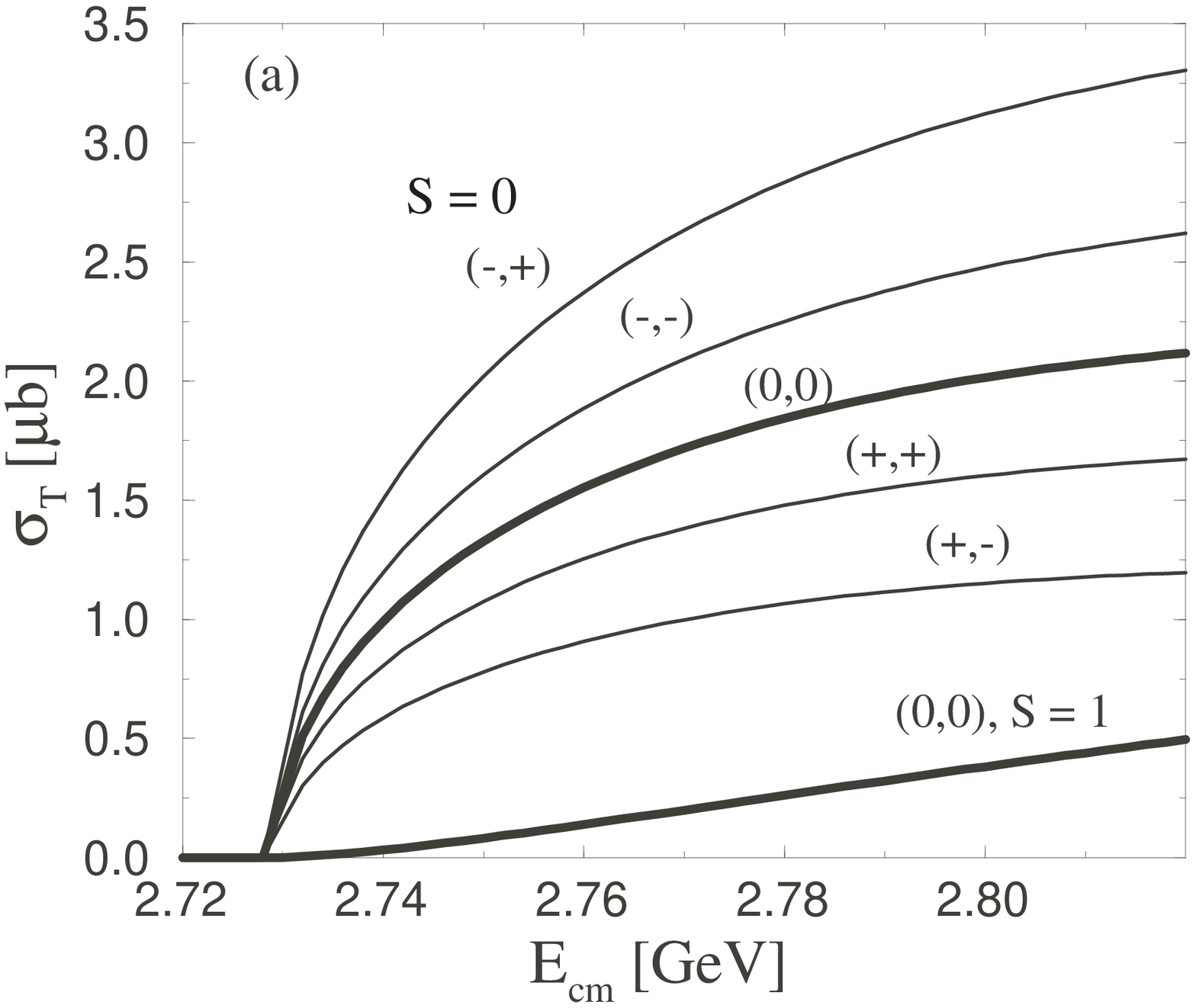}
\hspace*{0.5cm}
\includegraphics[width=6cm]{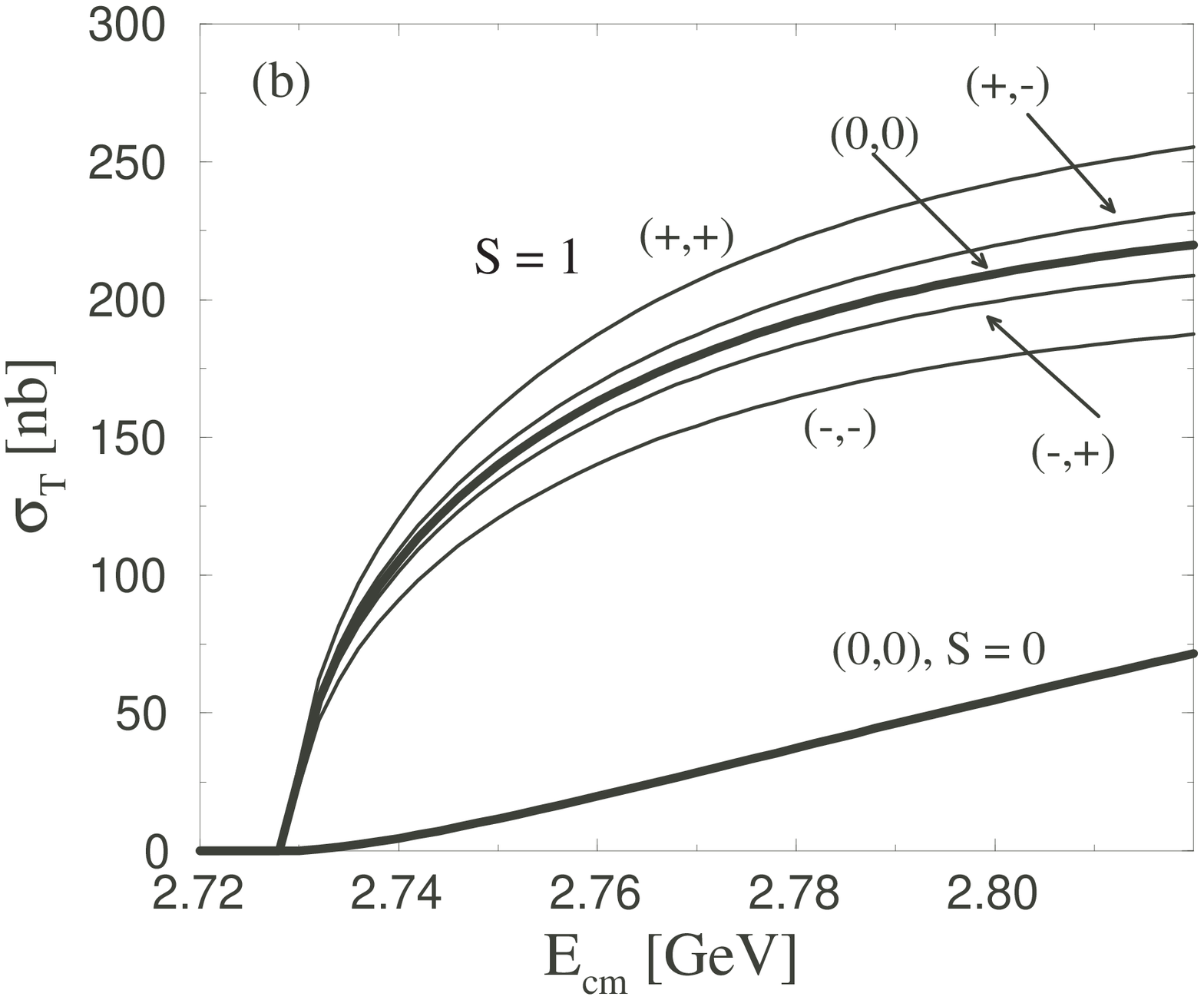}
\caption{Total cross sections near the threshold:   
(a) for positive parity $\Theta^+$ where the allowed channel 
is $S = 0$ and 
(b) for negative parity $\Theta^+$ where the allowed channel 
is $S = 1$.  
The labels (+,+) etc denote the signs of $g_{K^*N\Theta}$ and 
$g_{K^*N\Theta}^T$ relative to $g_{KN\Theta}$.  
The solid lines in the bottom is the cross sections 
for the forbidden channels.  
\label{stot}}
\end{center}
\end{figure}

The calculation for the scattering amplitude is straightforward once
having the interaction, Eqs.~(2) $\sim$ (5).
In Fig~\ref{stot}, total cross sections near threshold region are 
shown as functions of the energy in the center of mass system
$\sqrt{s}$ ($\sqrt{s}_{\rm th} = 2729.4$ MeV).  
The left (right) panel is for the positive (negative) parity 
$\Theta^+$ where the allowed initial spin is $S=0$ ($S=1$).
For the allowed channels, five curves are shown using different 
coupling constants of $g_{K^*N\Theta}$ and $g_{K^*N\Theta}^T$; 
zero and four
different combinations of signs with the absolute values 
$|g_{K^*N\Theta}^T| = 2|g_{K^*N\Theta}| = |g_{KN\Theta}|$, as indicated 
by the pair of labels in the figures, 
(sgn($g_{K^*N\Theta}$), sgn($g_{K^*N\Theta}^T)$).  
As shown in the figure, cross sections vary with about 50 \% 
from the mean value for the vanishing $K^*$ exchanges.   
For the forbidden channels only the case of vanishing $K^*N\Theta$ 
coupling constants is shown; cross sections using finite coupling constants 
vary within about 50 \% just as for the allowed channels.  
In both figures, the s-wave threshold behavior is seen for the 
allowed channels as proportional to $(s - s_{th})^{1/2}$, while the 
forbidden channels exhibit the p-wave dependence of $(s - s_{th})^{3/2}$ 
and with much smaller values than the allowed channel. 
The suppression factor is given roughly by 
[(wave number)$\cdot$(interaction range)]$^2$ 
$\sim k/m_K \sim 0.1$ ($k = \sqrt{2m_{K}E}$), as consistent with 
the results shown in the figures.  

From these results, we conclude that 
the absolute value of the total cross section is 
of the order 1 [$\mu$b] for the positive parity $\Theta^+$ and 
of the order 0.1 [$\mu$b] for the negative parity $\Theta^+$.  
The fact that the positive parity case has larger cross section 
is similar to what was observed in the photoproduction and hadron
induced reaction also~\cite{theory}.  
This is due to the p-wave nature of the $KN\Theta$ coupling with a 
relatively large momentum transfer for the $\Theta^+$ production.  
When the smaller decay width of $\Theta^+$ is used, the result
simply scales as proportional to the width, if the 
$K^*N\Theta$ couplings are scaled similarly.  

\begin{figure}[tbh]
\begin{center}
\includegraphics[width=6cm]{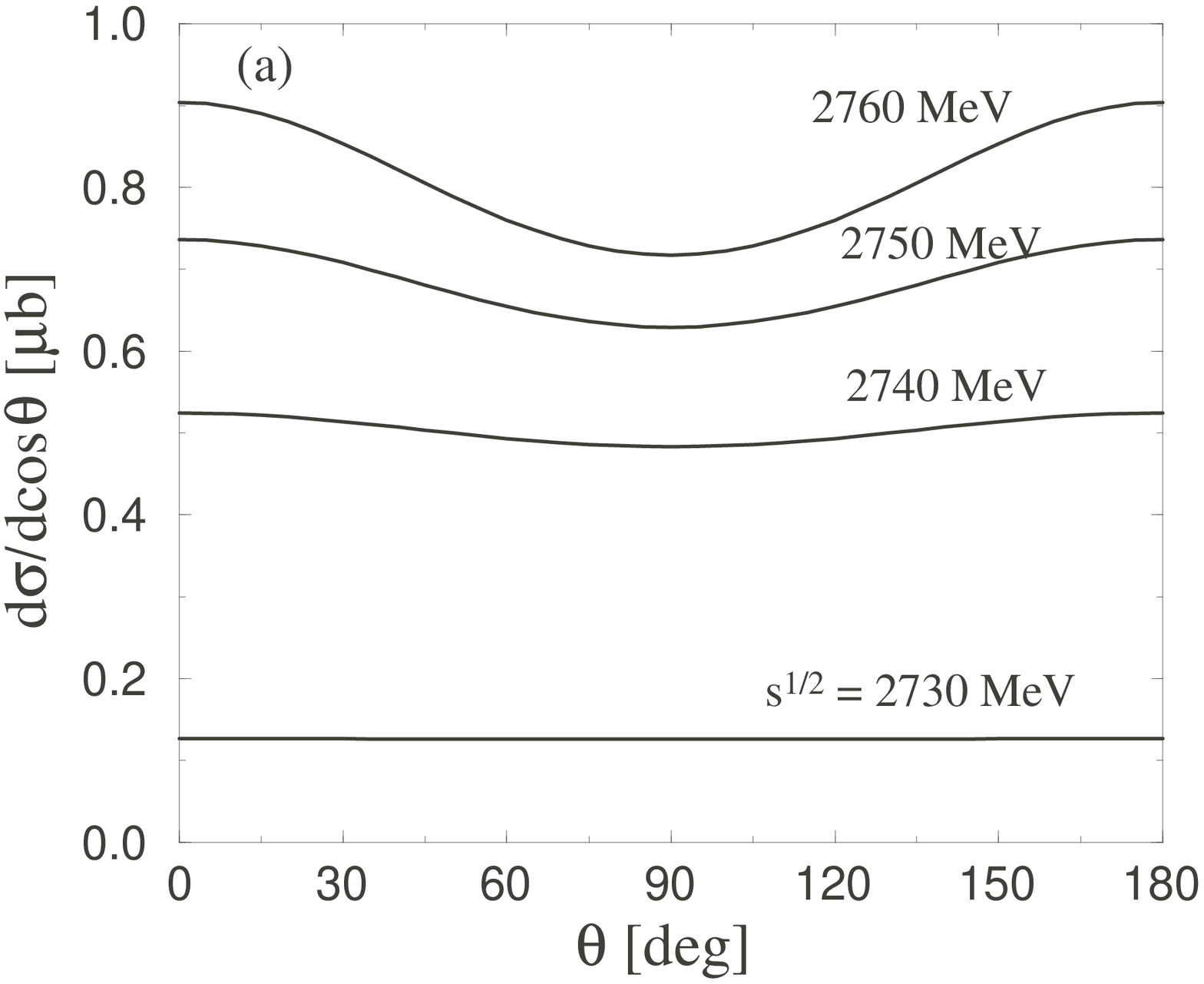}
\hspace*{0.5cm}
\includegraphics[width=6cm]{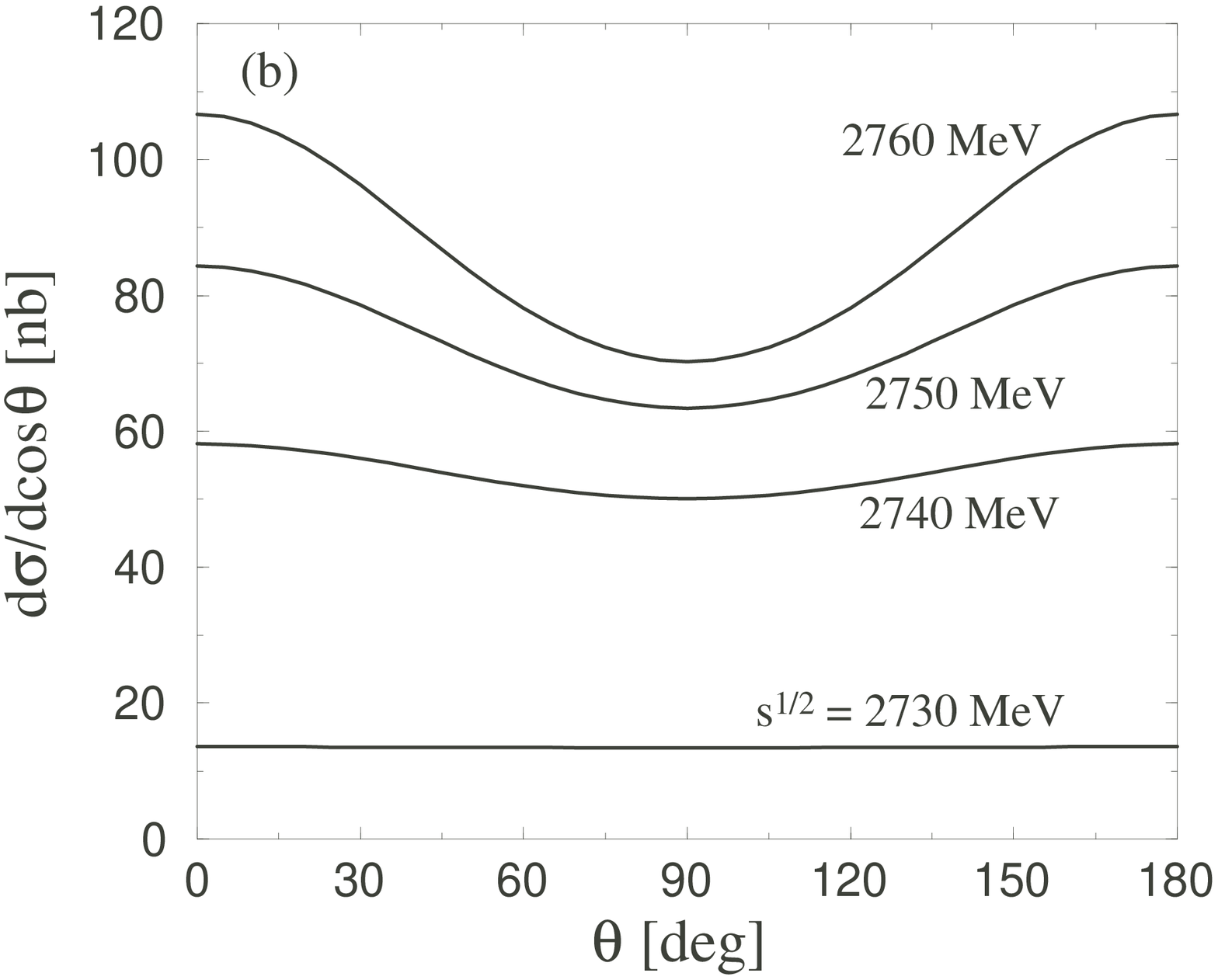}
\caption{Angular dependence of the production cross sections 
near the threshold in the center of mass frame:  
(a) for positive parity $\Theta^+$ and 
(b) for negative parity $\Theta^+$.  
The labels denote the total incident energy $\sqrt{s}$.  
\label{dsdt}}
\end{center}
\end{figure}

In Fig.~\ref{dsdt}, we show the angular dependence in the center 
of mass system for several different energies above the threshold, 
$\sqrt{s} =$ 2730, 2740, 2750 and 2760 MeV.  
Here only $K$ exchange is included but without $K^*$ exchanges. 
The angular dependence with the $K^*$ exchanges included is similar 
but with absolute values scaled as in the total cross sections.  
Once again, we can verify that the s-wave dominates the production 
reaction up to $\sqrt{s} \lsim$ 2750. 

\begin{figure}[tbh]
\begin{center}
\includegraphics[width=6.5cm]{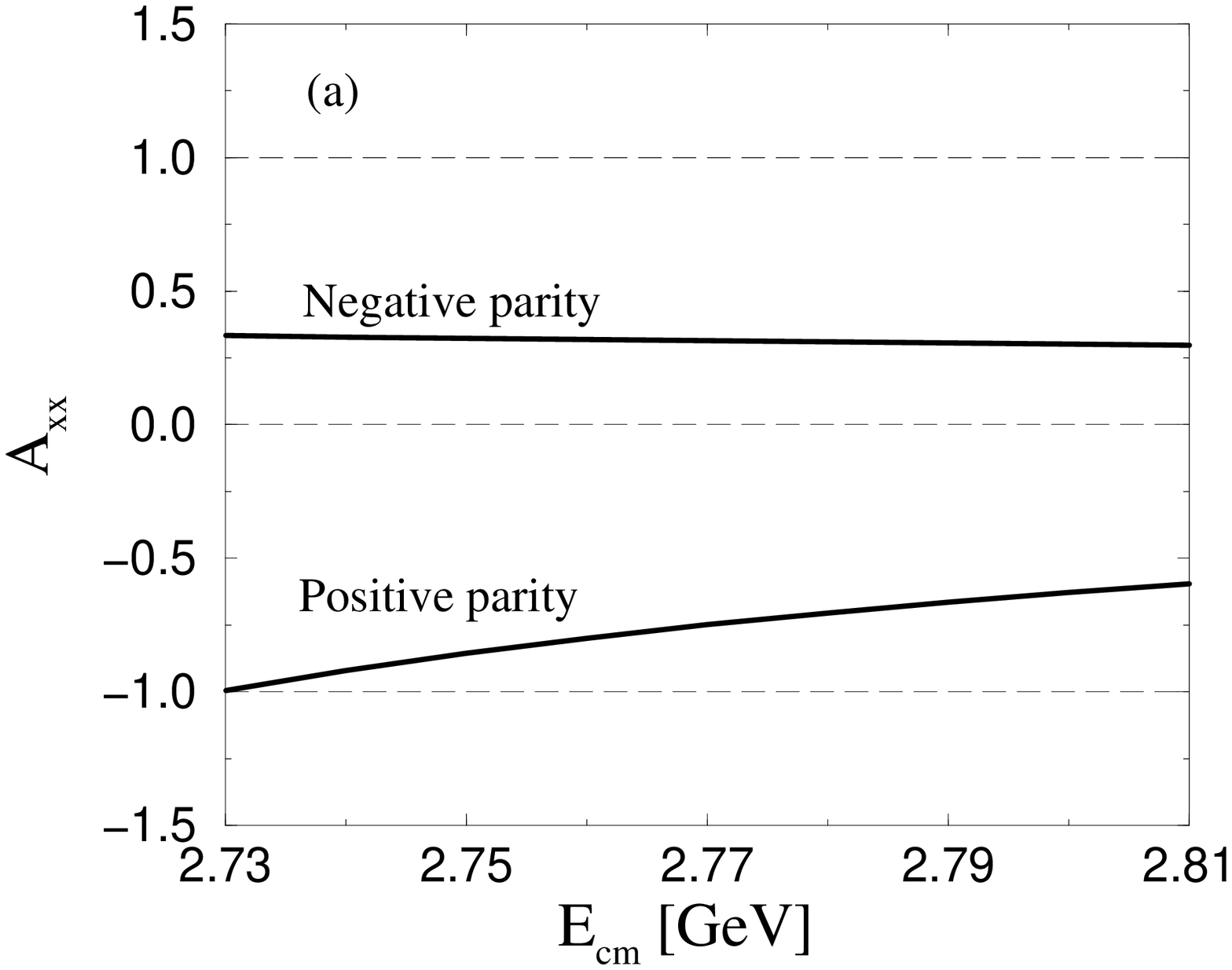}
\hspace*{0.5cm}
\includegraphics[width=6.5cm]{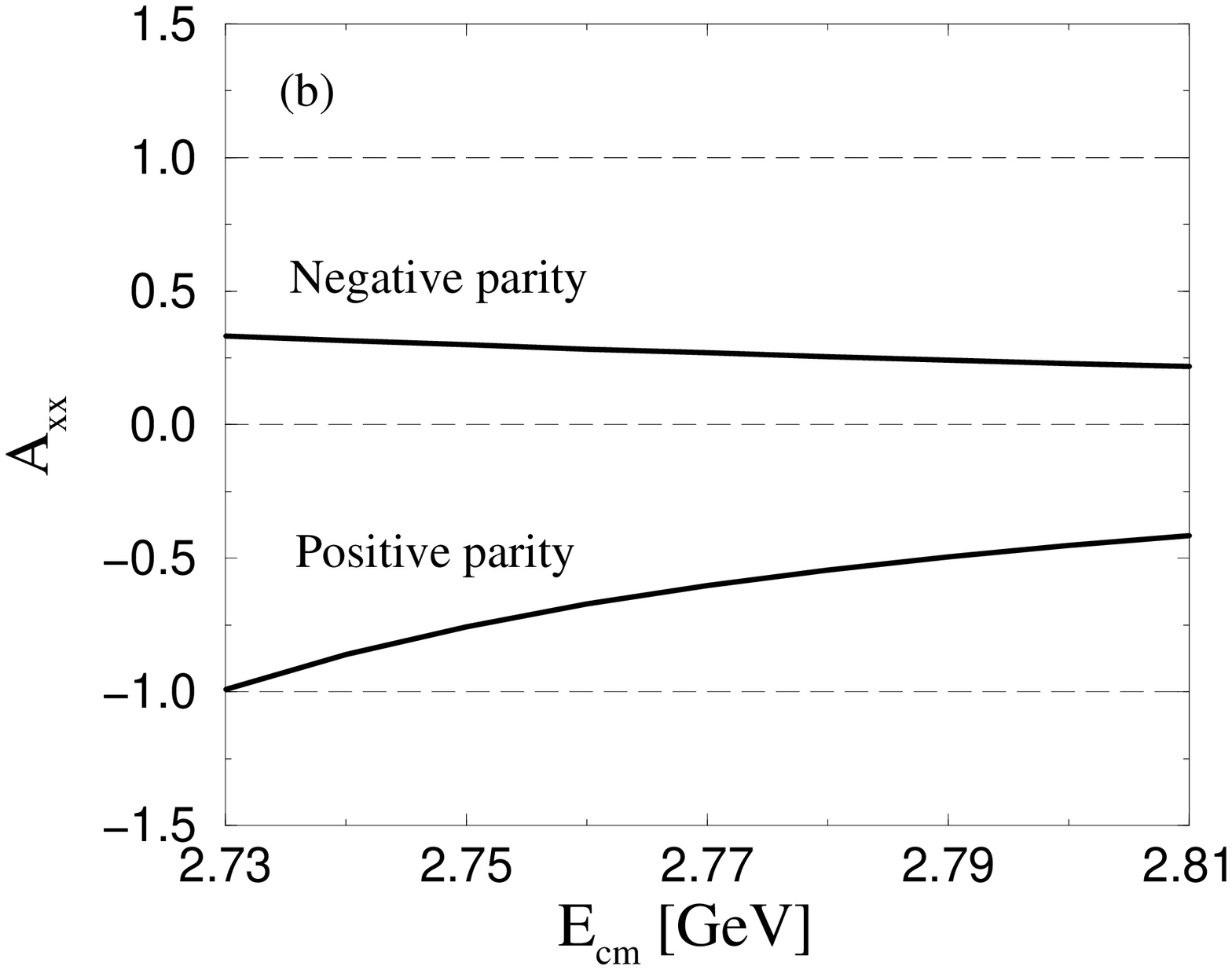}
\caption{$A_{xx}$ for the negative and positive parity of the $\Theta$
baryon. Here we do not consider $K^{*}$
exchange. (a) is without the form factor while (b) with. \label{axx}} 
\end{center}
\end{figure}

Recently, in Ref.~\cite{Hanhart:2003xp}, the authors discussed the
experimental methods and obervables to determine the parity of the
$\Theta^+$ baryon 
with the polarized proton beam and target. They discussed the spin
correlation parameter $A_{xx}$ as well as cross sections. It is computed by
\be
A_{xx}=\frac{(^{3}\sigma{_0}+^{3}\sigma_{1})}{2\sigma_{0}}-1,
\ee
where $\sigma_{0}$ is the unpolarized total cross sections and the
polarized cross section are denoted as $^{2S+1}\sigma_{S_z}$. In
Fig.~\ref{axx} we present $A_{xx}$ where we do 
not include $K^{*}$ exchange, but the results do not change very much
by including $K^*$. As shown in the figures $A_{xx}$ reflects very
clearly the differences of the parity of $\Theta^+$. The cases with and
without the form factor are similar and well fall into the region as
indicated in Ref.~\cite{Hanhart:2003xp}.   

In actual experiment, it is necessary to detect $\Sigma$ also at the 
threshold region.  
Due to small energy (or velocity) 
of the final state particles
in the center of system, produced $\Sigma$ must be detected inside
a very narrow cone forward peaked in the laboratory frame.  
Because of this fact, measurement at the existing facility of fixed 
target, such as COSY, would require an experimental challenge.

\section*{Acknowledgments}
We thank Tony Thomas, Ken Hicks, Hiroshi Toki, Kichiji Hatanaka and 
Takashi Nakano for discussions and comments. 
The work of H.-Ch.Kim is supported by the Korean Research Foundation
(KRF--2003--070--C00015). The work of S.I.Nam has been supported by
the scholarship endowed from the Ministry of Education, Science,
Sports and Culture of Japan.

\end{document}